\documentclass[aps,pra,superscriptaddress,preprint,showpacs,floatfix]{revtex4}

\usepackage{amsmath}
\usepackage{amssymb}
\usepackage{bm, color}
\usepackage{graphicx,epsfig,color}

\begin{document}


\title{On the resonant optical bistability condition.}

\author{A.\ V.\ Malyshev}
\email{a.malyshev@fis.ucm.es}

\affiliation{GISC, Departamento de F\'{\i}sica de Materiales, Universidad
Complutense, E-28040 Madrid, Spain}

\affiliation{on leave from: A. F. Ioffe Physical-Technical Institute, 194021
St.  Petersburg, Russia}

\date{\today}

\begin{abstract}

We address a two level system in an environment interacting with the
electromagnetic field in the dipole approximation.  The resonant optical
bistability induced by local-field effects is studied by considering the
relationship between the population difference and the excitation field. 
The diversity of various systems is included by accounting for the system
self-action via the surface part of the Green's dyadic in the general form. 
The bistability condition and the exact solution of the steady state optical
Bloch equations at the absolute bistability threshold are derived
analytically.

\end{abstract}


\maketitle


Because of its underlying nature and possible applications in the field of
all-optical processing, the optical bistability (OB) was a subject of an
intense experimental and theoretical research.\cite{Abraham82,Gibbs85} In
the early studies the use of a saturable absorber to induce OB was
suggested.\cite{Szoke69,Austin71,Spiller71,McCall74,Bonifacio76} The
phenomenon was demonstrated experimentally for a cell of sodium vapor
enclosed in a Fabry Perot interferometer and excited by a cw
dyelaser.\cite{Gibbs76} Later it was conjectured that the local-field
corrections alone could give rise to the mirrorless OB.~\cite{Bowden79} This
type of bistability was extensively
studied\cite{Abram82,Hopf84,BenAryeh86,Friedberg89,Crenshaw92} and observed
experimentally.\cite{Hehlen94}

The practical interest in all-optical devices faded to some extent as their
solid state counterparts proved to perform better in terms of the
switching speed and device density.  However, the OB and optical hysteresis
remain of considerable interest from the fundamental standpoint as a clear
manifestation of a nonlinear light-matter interaction.  Some new types of
bistability mechanisms were discussed
recently.\cite{Afanasev99,Kaplan08,Kaplan09,Volkov10} Besides, the question
of the OB and hysteresis has received a renewed attention in connection with
novel hybrid 0D nanoscopic systems, {\it e.  g.}, an artificial molecules
comprising a semiconductor quantum dot (SQD) and metal nanoparticles
(see Refs.~\onlinecite{Zhang06,Artuso08,Sadeghi10b,Malyshev11} and
references therein).

In this paper we address only the mirrorless OB induced by the local-field
effects on a two-level system (TLS) interacting with the electromagnetic
field in the dipole approximation.  The local-filed correction leads to a
self-action of the system, which results in a nonlinear relation between the
applied field and the one acting upon the system. This type of the OB
mechanism can be relevant for a large variety of systems: dense 3D
assemblies of two-level atoms~\cite{Bowden79,Friedberg89}, optically dense
thin films of TLS~\cite{Zakharov88,Benedict90} or films of linear molecular
aggregates,~\cite{Malyshev96,Klugkist07,Klugkist08} hybrid
metal-semiconductor systems,~\cite{Zhang06,Artuso08,Sadeghi10b,Malyshev11}
or a more general case of a TLS in an environment involving dielectric and
conducting surfaces, such as, a stratified media, a microcavity or a
nanostructure.

OB can occur within a range of internal system parameters and external field
intensities; identifying these ranges is therefore an important problems and
its analytical solution is desirable.  To the best of the author's
knowledge, so far, it has been solved exactly only for two particular cases. 
Thus, Friedberg {\it et al} obtained analytically the bistability condition
for the limiting case in which the active mechanism of the feedback is the
nonlinear Lorentz shift of the resonance in a 3D gas of two-level atoms,
which resulted from the near-field corrections.~\cite{Friedberg89} Ignoring
these corrections, Zakharov and Manykin derived the exact bistability
criterion in the other limit in which the self-action is due to the radiated
secondary field in a thin 2D film of TLS.~\cite{Zakharov88} The case when
both these fields contribute into the self-action has been studied only
numerically so far.~\cite{Orayevsky94}

We consider the simplest optical TLS coupled to an environment.  The
self-action of the TLS due to the environment can be described by the
surface part of the Green's dyadic evaluated at the position of the TLS. 
Therefore, in this case the nonlinearity is characterized by a complex
number which should satisfy some condition for the OB to occur.  Below, we
derive such condition analytically.

Following Refs.~\onlinecite{Friedberg89,Zakharov88,Orayevsky94} we address
the OB condition by considering the relationship between the population
difference and the excitation field within the framework of the Bloch
equations for the $2\times 2$ density matrix $\rho_{mn} \> (m,n = 0,1)$ of a
TLS.  In the rotating wave approximation these equations read:
\begin{subequations}
\begin{align}
\label{dotZ} \dot Z & = - \gamma\,(Z+1)-\frac{1}{2} \left[\,
\Omega\,P^{*}+\Omega^{*}P \,\right]\ ,\\
\label{dotP} \dot P & = -(\Gamma+i\,\Delta)\, P+\Omega\,Z \ ,
\end{align}
\label{MB}
\end{subequations}
where $Z=\rho_{11}-\rho_{00}$ is the population difference between the
excited and the ground state of the TLS, $P$ is the amplitude of the
off-diagonal density matrix element defined through $\rho_{10} =
-(i/2)P \exp(-i\omega t)$, $\gamma$ and $\Gamma$ are the relaxation
constants of the population and the dipole moment, respectively, $\Delta =
\omega_0 - \omega$ is the detuning of the driving field frequency $\omega$
from the TLS resonance $\omega_0$, and $\Omega={\bm\mu\bf E}/\hbar$ is the
total electric field $\bf E$ (in frequency units) acting upon the system,
while $\bm\mu$ being the TLS optical transition dipole.  The dipole moment
of the system is ${\bf p}=-i\,P{\bm\mu}$, where $P$ is its complex
amplitude.

To calculate the total field $\Omega$ the Maxwell's equations are to be
solved for a particular geometry.  However, for a single dipole in an
environment the total field $\Omega$ can be represented in the following
general form:
\begin{equation}
\label{Omega}
\Omega = \widetilde\Omega_0 +\Omega_2,\quad \Omega_2 = - i\,G\,P
\end{equation}  
where $\widetilde\Omega_0$ is the renormalized incident field and $\Omega_2$
is the secondary dipole field.  The latter field is due to the feedback of
the environment and is related to the corresponding response
function\cite{Agarwal75a,Agarwal75b,Agarwal98}.  It is determined by the
surface part of the Green's dyadic evaluated at the position of the dipole. 
The near-zone component of the secondary dipole field, governed by
$\mathrm{Re}(G)$, can originate from the near-field
corrections,~\cite{Friedberg89} while the far-zone contribution, governed by
$\mathrm{Im}(G)$, can be due to the radiation.~\cite{Zakharov88} In hybrid
nano-systems comprising a SQD, the feedback is provided by the secondary
reflected field of the optical transition dipole moment.  Thus,
independently of its physical mechanism, the self-action is determined by
the single complex valued parameter $G$.  The total field $\Omega$ acting
upon the TLS depends therefore on its state.  Together with the nonlinearity
of the TLS that can give rise to a variety of effects, such as the OB.

\begin{figure}[t!]
\begin{center}
\includegraphics[width=0.9\columnwidth]{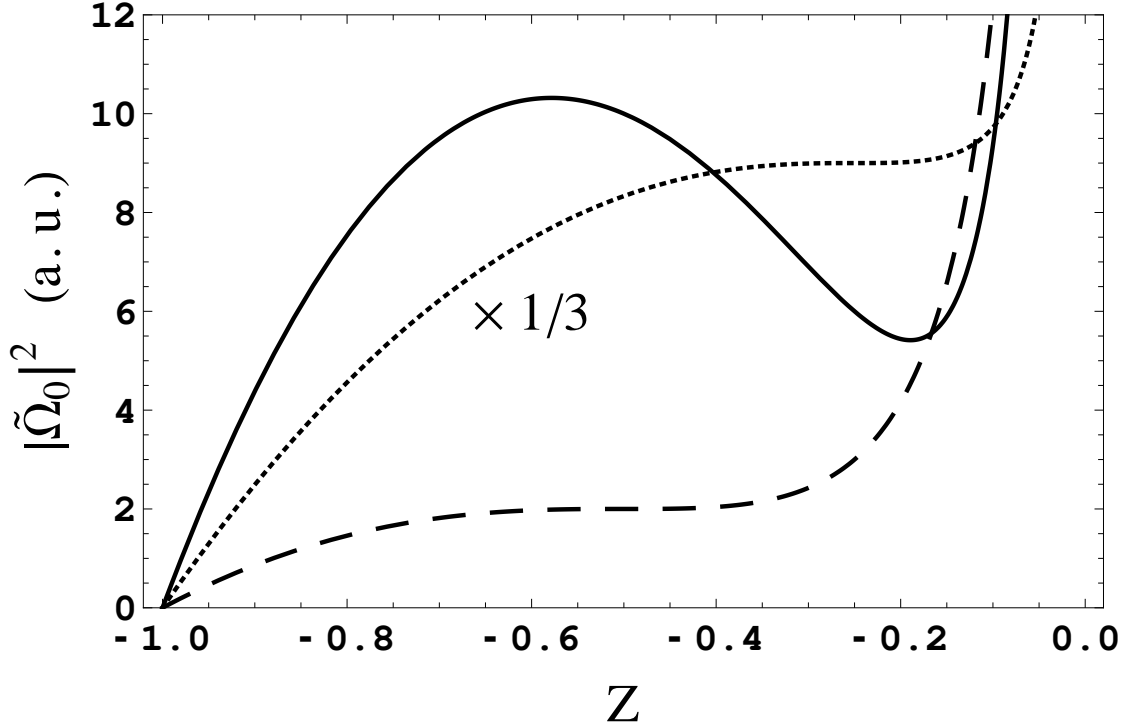}
\end{center}
\caption{The RHS of Eq.~(\ref{Z}) for $G=6$, $\Delta=1$ (solid line),
$G=4$, $\Delta=1$ (dashed line) and $G=8\,i$, $\Delta=0$ (dotted line)}
\label{OmegaOnZ}
\end{figure}
Under the steady-state conditions Eqs.~(\ref{MB}) read:
\begin{subequations}
\label{stat}
\begin{align}
\label{Z}
    \frac{|\widetilde\Omega_0|^2}{\gamma\,\Gamma} & =
    -\frac{Z+1}{Z}\> \frac{(\Gamma - G_\mathrm{I}Z)^2
	+ (\Delta + G_\mathrm{R}Z)^2} {\Gamma^2}\ ,\\
\label{P}
    P & = \frac{Z\;\widetilde\Omega_0}
	{(\Gamma - G_\mathrm{I}\,Z)+i\,(\Delta + G_\mathrm{R}\,Z)}\ ,
\end{align}
\end{subequations}
where $G_\mathrm{R} = \mathrm{Re} (G)$ and $G_\mathrm{I} = \mathrm{Im} (G)$. 
The above equation highlights a known result of cavity quantum
electrodynamics: the real part of the Green's dyadic shift the frequency
while the imaginary part renormalizes the decay
rate.\cite{Agarwal75a,Agarwal75b,Agarwal98} Both effects can be active
mechanisms of the OB.

Eq.~(\ref{Z}) is of the third order in $Z$ and can therefore have three real
roots for some values of $\Delta$, $\Gamma$, $G_\mathrm{R}$, and
$G_\mathrm{I}$ ($\Gamma$ is used as a unit rate from now on).  These three
solutions are different when the right hand side (RHS) of Eq.~(\ref{Z}) has
a minimum and a maximum (see the solid line in Fig.~\ref{OmegaOnZ}),
corresponding to two real roots of the RHS derivative, which satisfy the
following equation
\begin{equation}
2\,|G|^2\,Z^3+(|G|^2+2\,G_\mathrm{R}\,\Delta-2\,G_\mathrm{I})\,Z^2-(1+\Delta^2)=0\ .
\label{RHS}
\end{equation}
A threshold for bistability occurs when Eq.~(\ref{RHS}) has a double root
(merged extrema -- see the dashed or dotted lines in Fig.~\ref{OmegaOnZ}). 
In this case the solutions of Eq.~(\ref{Z}) are:
\begin{equation}
Z_1=Z_2=-\frac{|G|^2+2\,(G_\mathrm{R}\Delta-G_\mathrm{I})}{3\;|G|^2}\ ,\quad 
Z_3 = -\frac{Z_1}{2}
\label{roots}
\end{equation}
Using the above root $Z_{1,2}$ and the condition $-1 \leq Z \leq 0$ one
obtains an important constraint for the detuning $\Delta$:
\begin{equation}
G_\mathrm{R}\,\Delta - G_\mathrm{I} \leq |G|^2
\label{constr}
\end{equation}
The formal condition of the existence of the double root (\ref{roots})
is determined by the following equation on $\Delta$:
\begin{equation}
\left( |G|^2+2\,G_\mathrm{R}\,\Delta-2\,G_\mathrm{I} \right)^3
- 27\,|G|^4\,(1+\Delta^2) = 0\ ,
\label{Delta}
\end{equation}
which is cubic in $\Delta$ and can also have three real solutions. It can be
shown that only two of them, say $\Delta_1<\Delta_2$, satisfy the constraint
(\ref{constr}) and yield the upper and lower threshold values of the
population difference $Z$ and the external field $\widetilde\Omega_0$. 
Therefore, Eq.~(\ref{RHS}) has three different real roots within a window of
detunings $\Delta_1\leq\Delta\leq\Delta_2$.  The {\it absolute} bistability
threshold occurs when the window shrinks into a point, {\it i.  e.}, when a
degenerate root $\Delta_1=\Delta_2$ of Eq.~(\ref{Delta}) appears.  After
some algebra, one can obtain the following alternative for the existence of
such double root:
\begin{subequations}
\begin{align}
\label{GR}
G_\mathrm{R}^4+2\,(G_\mathrm{I}^2-4\,G_\mathrm{I}-8)\,G_\mathrm{R}^2
+G_\mathrm{I}^3\,(G_\mathrm{I}-8) & = 0\ ,\\
G_\mathrm{I}+1 & = 0\ .
\label{GI}
\end{align}
\label{equ}
\end{subequations}
\begin{figure}[t!]
\begin{center}
\includegraphics[width=0.7\columnwidth]{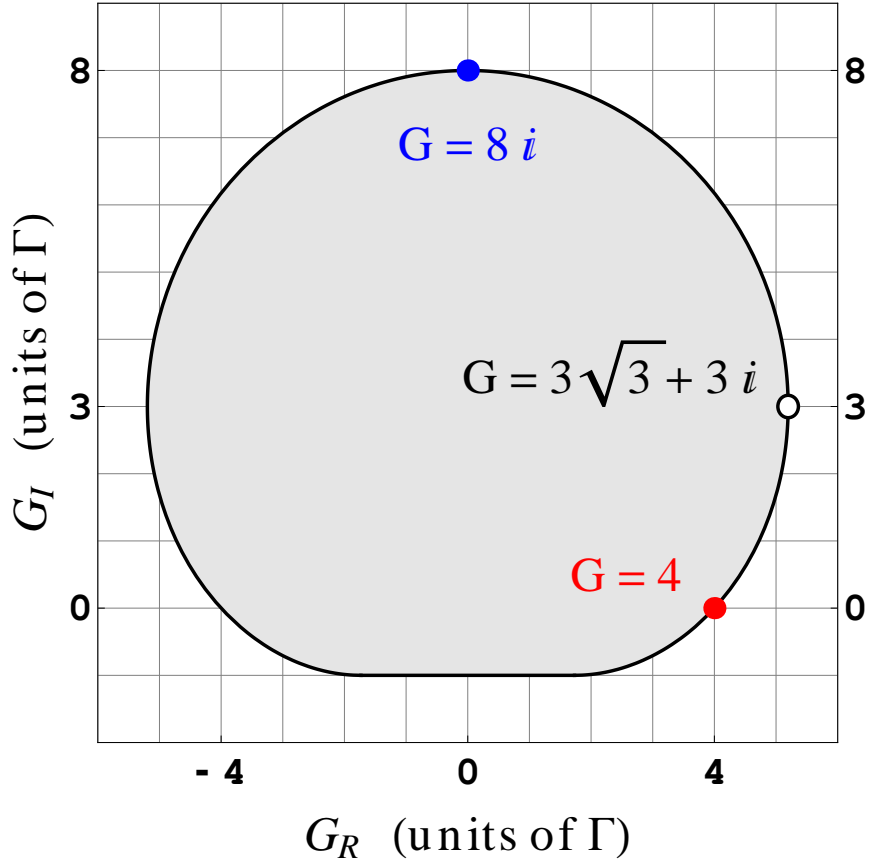}
\end{center} 
\caption{The stability phase map: outside the gray area the system can be
bistable.  Solid line represents the phase boundary given by
Eq.~(\ref{crit}). See text for more details.}
\label{phase}
\end{figure} 

Only two of the solutions of the biquadratic equation~(\ref{GR}) satisfy the
constraint (\ref{constr}).  Analyzing the roots of Eq.~(\ref{RHS}) in the
vicinity of the solution to Eqs.~(\ref{GI}) and (\ref{Delta}), one finds
that the system can be bistable below the line given by $G_\mathrm{I}=-1$,
while being critical ({\it i.  e.}, $Z_{1,2}=-1$) on the line for
$|G_\mathrm{R}|\leq\sqrt{3}$.  Finally, the absolute bistability threshold
is:
\begin{subequations}
\label{crit}
\begin{align}
G_\mathrm{R}^2 = 8+4\,G_\mathrm{I}-G_\mathrm{I}^2+8\sqrt{1+G_\mathrm{I}},& \; -1 < G_\mathrm{I} \leq 8\\
-\sqrt{3} \leq G_\mathrm{R} \leq \sqrt{3},&  \qquad \;\; G_\mathrm{I}=-1
\end{align}
\end{subequations}
For completeness, we provide also the expression for the corresponding
detuning:
\begin{equation}
\label{delta}
\Delta=\frac{|G|^4-4\,G_\mathrm{R}^2\,(G_\mathrm{I}+3)
-4\,G_\mathrm{I}^2\,(G_\mathrm{I}-1)}
{|G|^4+8\,G_\mathrm{I}\,\left[\,G_\mathrm{I}^3
+G_\mathrm{R}^2\,(G_\mathrm{I}+2)\,\right]}\;G_\mathrm{R}\ .
\end{equation}

Equation (\ref{crit}) provides the condition of existence of three real
roots of Eq.~(\ref{Z}).  We studied the stability of these by analyzing
Lyapunov exponents of Eqs.~(\ref{MB}) in the vicinity of a stationary
solution and calculated the stability phase map numerically. 
Figure~\ref{phase} shows such phase diagram in the space of the feedback
parameter $G$.  Three different real solutions of Eq.~(\ref{Z}) exist within
the white area, two of them being stable.  The line dividing the two phases
represents the analytical condition Eq.~(\ref{crit}) which gives also the
bistability threshold.  The general criterion Eq.~(\ref{crit}) recovers both
reported exact results: $G_\mathrm{I}>8$ for
$G_\mathrm{R}=0$~\cite{Zakharov88} and $G_\mathrm{R}>4$ for
$G_\mathrm{I}=0$~\cite{Friedberg89} (the corresponding threshold values of
$G$ are marked by full circles in Fig.~\ref{phase}).  Note the mirror
symmetry of the bistability map with respect to the line $G_\mathrm{R}=0$,
which reflects the invariance of Eq.~(\ref{Z}) under the simultaneous change
of signs of $G_\mathrm{R}$ and $\Delta$~\cite{Orayevsky94}.

Eqs.~(\ref{roots}) and (\ref{delta}) give the full solution of the steady
state problem at a bistability threshold, while Eq.~(\ref{crit}) is the
bistability condition, which constitute the main result of the paper.

Finally, we note that novel hybrid plasmonic nano systems, such as a SQD and
a metal nanoparticle complex (see Ref.~\onlinecite{Malyshev11} and
references therein) or a SQD embedded in a stratified medium are excellent
model systems with adjustable nonlinearity.  The self-action field in these
complexes is the secondary reflected field of the SQD optical transition
dipole moment acting back upon the SQD.  In the most general case such
environment feedback results in a complex self-action parameter $G$ which
can be engineered by an appropriate choice of materials, geometry and/or
external control parameters.  Thus, if the SQD transition frequency is far
from the plasmon resonance of the system then, typically, $|G_\mathrm{R}|
\gg |G_\mathrm{I}|$ and the dominant mechanism of the bistability is the
nonlinear Lorentz shift~\cite{Friedberg89}.  If the excitation frequency is
near the surface plasmon resonance then the feedback can be almost purely
imaginary: $|G_\mathrm{I}| \gg |G_\mathrm{R}|$ with $G_\mathrm{I}>0$. 
Algebraically, this case is equivalent to the one of the radiation induced
self-action~\cite{Zakharov88}.  A very interesting case is $G_\mathrm{I}<0$
which can hardly be realized in the traditional extended 3D or 2D systems of
two-level atoms.  Nanoscopic hybrids are much more promising from this point
of view due to their greater diversity.  In the latter case various
instabilities such as auto-oscillations can be expected under steady-state
excitation.  However, the detailed study of these instabilities goes beyond
the scope of the present contribution and will be analyzed elsewhere.

Summarizing, we addressed the mirrorless resonant optical bistability of a
two-level system in an environment by considering the relationship between
the population difference and the intensity of the excitation field.  The
feedback of the environment was introduced in a general from via the surface
part of the Green's dyadic.  We derived the analytical bistability condition
and obtained the exact steady state solution of the Maxwell-Bloch equations
at the absolute bistability threshold.  Our findings open the possibility to
easily analyze diverse physical systems, determine experimental conditions
(such as, the appropriate detuning from the bare resonance and the intensity
of the external field) necessary to observe the bistability and other
nonlinear optical effects, as well as design and engineer new systems with
desirable nonlinear optical properties.

\section{Acknowledgments}

Support from the project MOSAICO (FIS2006-01485), fruitful discussions with
V.  A.  Malyshev and the hospitality of the University of Groningen are
gratefully acknowledged.

\end{document}